\documentclass[english,aip,rsi,reprint,graphicx]{revtex4-1} 

\usepackage{graphicx}
\usepackage{dcolumn}
\usepackage{bm}
\usepackage[LGR,T1]{fontenc}
\usepackage[latin9]{inputenc}
\usepackage{textcomp}
\makeatletter

\DeclareRobustCommand{\greektext}{%
  \fontencoding{LGR}\selectfont\def\encodingdefault{LGR}}
\DeclareRobustCommand{\textgreek}[1]{\leavevmode{\greektext #1}}
\DeclareFontEncoding{LGR}{}{}
\DeclareTextSymbol{\~}{LGR}{126}

\makeatother

\usepackage{babel}

\draft 

\begin{document}


\title{A Novel Femtosecond-Gated, High-Resolution, Frequency-Shifted Shearing
Interferometry Technique for Probing Pre-Plasma Expansion in Ultra-Intense
Laser Experiments} 

\thanks{Contributed paper published as part of the Proceedings of the 20th Topical Conference on High-Temperature Plasma Diagnostics, Atlanta, Georgia, June, 2014.\\}


\author{S. Feister}
 \email{feister.7@osu.edu}
 \affiliation{Department of Physics, The Ohio State University, Columbus, OH, USA}
 \affiliation{Innovative Scientific Solutions, Inc., Dayton, OH, USA}

\author{J.A. Nees}%
 \affiliation{\mbox{Center for Ultra-Fast Optical Science, University of Michigan, Ann Arbor, MI, USA}} 
 \affiliation{Innovative Scientific Solutions, Inc., Dayton, OH, USA}

\author{J.T. Morrison}%
 \affiliation{Fellow, National Research Council, USA}

\author{K.D. Frische}%
 \affiliation{Innovative Scientific Solutions, Inc., Dayton, OH, USA}

\author{C. Orban}%
 \affiliation{Department of Physics, The Ohio State University, Columbus, OH, USA}
 \affiliation{Innovative Scientific Solutions, Inc., Dayton, OH, USA}

\author{ E.A. Chowdhury}%
 \affiliation{Department of Physics, The Ohio State University, Columbus, OH, USA}
 \affiliation{Intense Energy Solutions, LLC., Plain City, OH, USA}

\author{W.M. Roquemore}
 \affiliation{Air Force Research Laboratory, Dayton, OH, USA}


%
%
%

\date{\today}

\begin{abstract}
Ultra-intense laser-matter interaction experiments (>10\textsuperscript{18}~W/cm\textsuperscript{2}) with dense targets are highly sensitive
to the effect of laser \textquotedblleft{}noise\textquotedblright{}
(in the form of pre-pulses) preceding the main ultra-intense pulse.
These system-dependent pre-pulses in the nanosecond and/or picosecond
regimes are often intense enough to modify the target significantly
by ionizing and forming a plasma layer in front of the target before
the arrival of the main pulse. Time resolved interferometry offers
a robust way to characterize the expanding plasma during this period.
We have developed a novel pump-probe interferometry system for an
ultra-intense laser experiment that uses two short-pulse amplifiers
synchronized by one ultra-fast seed oscillator to achieve 40-femtosecond
time resolution over hundreds of nanoseconds, using a variable delay
line and other techniques. The first of these amplifiers acts as the
pump and delivers maximal energy to the interaction region. The second
amplifier is frequency shifted and then frequency doubled to generate
the femtosecond probe pulse. After passing through the laser-target
interaction region, the probe pulse is split and recombined in a laterally
sheared Michelson interferometer. Importantly, the frequency shift
in the probe allows strong plasma self-emission at the second harmonic
of the pump to be filtered out, allowing plasma expansion near the
critical surface and elsewhere to be clearly visible in the interferograms.
To aid in the reconstruction of phase dependent imagery from fringe
shifts, three separate 120$^\circ$ phase-shifted (temporally
sheared) interferograms are acquired for each probe delay. Three-phase
reconstructions of the electron densities are then inferred by Abel
inversion. This interferometric system delivers precise measurements
of pre-plasma expansion that can identify the condition of the target
at the moment that the ultra-intense pulse arrives. Such measurements
are indispensable for correlating laser pre-pulse measurements with
instantaneous plasma profiles and for enabling realistic Particle-in-Cell
simulations of the ultra-intense laser-matter interaction.
\end{abstract}


\maketitle 


\section{Introduction}

In ultra-intense laser systems, so-called ``pre-pulses'' are inexorably
produced during the laser amplification process, and can pre-ablate
the target, changing the condition of the target at the time of arrival
of the main ultra-intense laser pulse. The outcome of this main pulse
interaction sensitively depends on these pre-ablations \cite{hosokai_effect_2003, kaluza_influence_2004, zhidkov_prepulse_2000, krygier_origin_2014}.
The pre-ablation of the target by the pre-pulse can be difficult to
predict \textit{ab initio} using hydrodynamic and particle-in-cell
codes in large part due to the wide range of timescales involved in
the laser-matter interaction. Relevant pre-pulses occur and their
effects evolve on the nanosecond and/or picosecond scale, with consequences
to the main laser-plasma interaction occurring on the femtosecond scale
\cite{hong_generation_2005, ivanov_amplified_2003}. Characterizing
and controlling these pre-plasma conditions is an integral part of
ultra-short pulse experiments \cite{orban_radiation-hydrodynamics_2013,dollar_scaling_2013}.
Interferometry is a well-known technique that can reveal the plasma
electron density profile \cite{hutchinson_principles_2005}. However,
the temporal resolution of interferometric measurements, even with
state-of-the-art streak camera technology, is typically limited to
picosecond timescales and these devices are prohibitively expensive.

We present a pump-probe technique with precision timing features that
allow interferometric phase reconstruction to occur on timescales
of less than one hundred femtoseconds, in a cost-effective manner,
while retaining roughly nine orders of magnitude temporal dynamic
range. Femtosecond timing stability between pump and probe beams is achieved through use
of a common oscillator. The probe
light is used in two ways: shadowgraphy reveals general features \cite{wu_time-resolved_2011},
while interferometry and an Abel inversion \cite{hutchinson_principles_2005}
recover phase and reveal subtle features of plasma evolution \cite{temnov_femtosecond_2004}. 

The contamination of shadowgraphs and interferograms by the plasma
self-emission (e.g. \cite{le_pape_characterization_2009,mckenna_effects_2008,dunn_recent_2001})
is avoided by selective optical filtering and a frequency shift of
the probe beam before amplification. These techniques create a unique
platform for performing spatially and temporally resolved measurements
of the target evolution. We show results from an experiment at the
Air Force Research Lab (AFRL) in which a flowing water jet target
\cite{uhlig_laser_2011,fullagar_broadband_2007} is irradiated by
nanosecond-scale and picosecond-scale pre-pulses that precede a 30-fs
FWHM ultra-intense (10\textsuperscript{18}~W/cm\textsuperscript{2})
interaction. Section II describes the experimental setup, and Section
III describes how the ultra-short timescale synchronization of the
probe and pump pulse is achieved. Section IV describes the frequency
shift of the probe pulse and selective filtering. Section V presents
some preliminary data of the water jet expansion due to laser-target
interactions, including Abel inversion of interferometric data. Section
VI states our conclusions and describes how this instrument will be
incorporated into future work.

\section{Experimental Overview}

\begin{figure*} 
\includegraphics[width=6in]{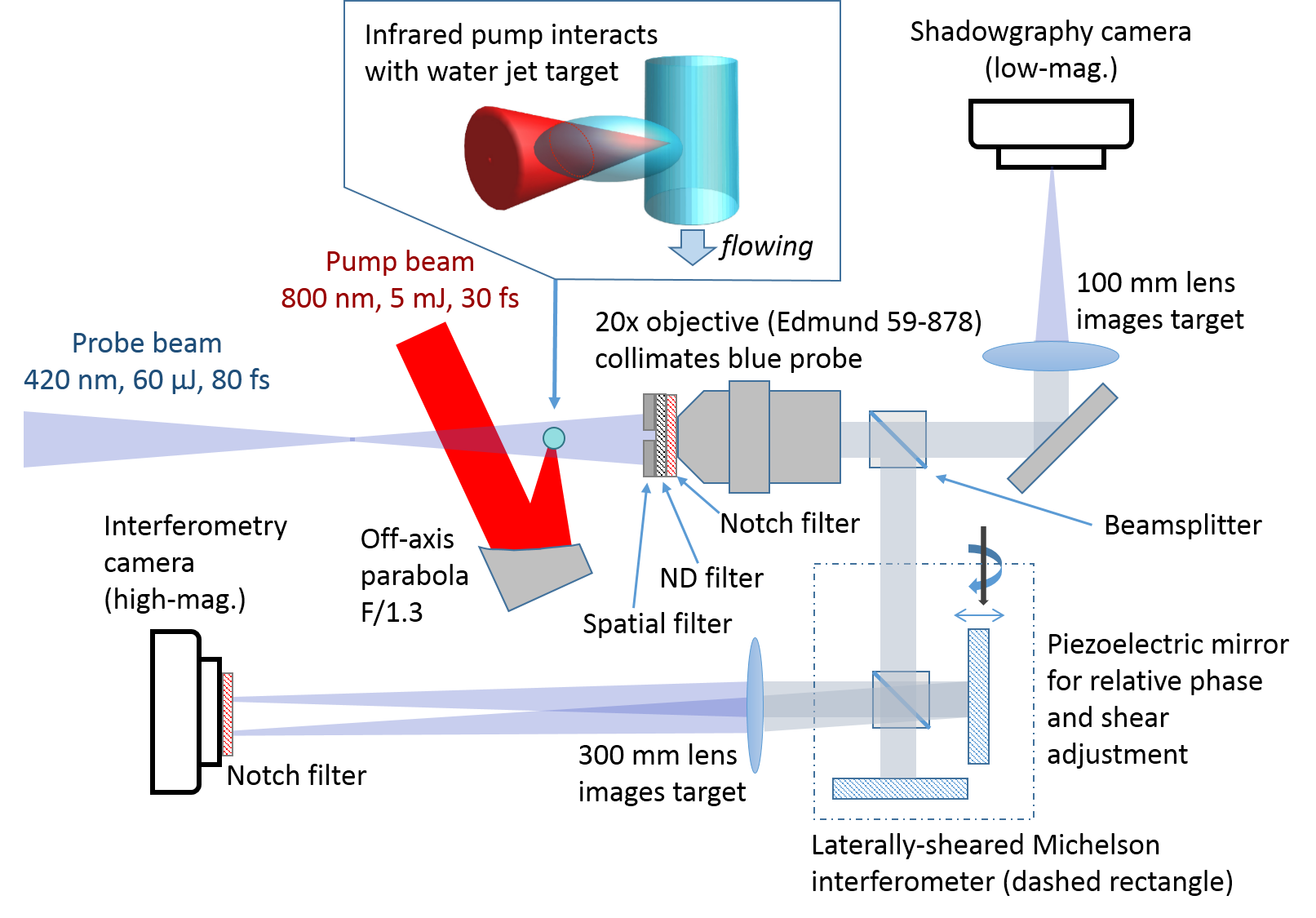}\caption{\label{fig:layout}The experiment employs a high-intensity pump and
low-intensity probe, relatively timed with better than 40-femtosecond
precision. The pump beam (sketched in red) and its pre-pulses irradiate
the water jet target, creating a dynamic laser-plasma interaction
region. The probe beam (sketched in blue) envelops the interaction
region and is then split for target-imaged interferometry and shadowgraphy. Within the laterally-sheared Michelson interferometer, a piezoelectric
mirror allows for controlled adjustment of phase (mirror translation)
and shear (mirror tilt). The shear angle is out of the page, though
sketched in the plane of the page for easy visualization. Optical notch filters (narrowband optical filters, Semrock FF01-420/10-25) and an iris (spatial filter) exclude unwanted light such as that from plasma self-emission
(see Fig. \ref{fig:notch}). }
\end{figure*}

Interferometry and shadowgraphy are used to characterize plasma expansion
in experiments at the Air Force Research Laboratory (AFRL) at Wright
Patterson Air Force Base in Dayton, OH. An 800~nm, 30-fs FWHM pump
beam produces high intensities (10\textsuperscript{18}~W/cm\textsuperscript{2},
2.6~\textgreek{m}m FWHM spot size) on flowing water jet column targets.
To avoid disruption of the high intensity laser light as it propagates
to focus and to prevent freezing of the water jet nozzle, the experiment
is housed in a vacuum chamber that is held at 20~Torr partial vacuum
using a thermocouple gauge solenoid valve feedback loop.

Known pre-pulse artifacts of the amplification process pre-ablate
the target, creating conditions of interest picoseconds and nanoseconds
before the main laser pulse interaction. To interrogate these conditions,
a 420~nm, 80-fs FWHM probe beam is passed through the interaction
region and subsequently split to image the target for shadowgraphy
and interferometry. The pump-probe experimental layout is shown in
Fig. \ref{fig:layout}. The probe beam is frequency-shifted, as discussed
in Section IV.

Distortions to the probe pulse wavefront due to variations in the
index of refraction along the line of sight are revealed by interfering
the probe pulse with a reference pulse at the same frequency. In this
laterally-sheared Michelson interferometer setup \cite{hung_speckle-shearing_1974},
an image of the interaction region of the water jet is overlapped
with a sheared image of the water jet downstream. Interference fringes
shift in response to changes in line-of-sight index of refraction, allowing one to ``see''
ablated liquid and ionized plasma in a way complementary to shadowgraphy.
With a high dynamic range timing setup (described next) and selective
optical filtering, one can get a sense of the evolution of this ablated
liquid and plasma.

\section{Femtosecond Resolution Over Microseconds}

One commonly-used approach to pump/probe experiment is to use two
entirely separate femtosecond laser systems, synchronizing the laser
oscillators via electronic signals \cite{crooker_femtosecond_1996}.
A typical setup gives picosecond stability between oscillators; achieving
femtosecond stability requires advanced electronics and a reduction
in the output repetition rate \cite{ma_sub-10-femtosecond_2001}.
Utilizing two pulses from a single oscillator is simpler and results in
a more stable relative timing.

To achieve femtosecond relative precision in such a setup, the probe
and pump beams must be seeded from a common oscillator. In a kHz system,
a Pockels cell selects one oscillator pulse every millisecond to be
further amplified as the pump pulse, rejecting all other pulses. The oscillator used in the experiment
at AFRL produces pulses at 80~MHz, and the rejected pulses
are routed into a second Pockels cell that selects a different pulse
to be amplified as the probe pulse. By varying the pulse selected, coarse delays can be introduced between the pump and
probe on the order of the oscillator rate, 80~MHz or 12.5~ns.
After amplification, the probe pulse passes through a delay line. The double-passed delay line allows fine adjustments
to the relative pump-probe delay, with <~40~fs resolution over 19~ns. Combining coarse and fine delay techniques results in 40~fs resolution over 10~\textgreek{m}s, and potentially even longer times
(e.g. 900~\textgreek{m}s) if desired. The final resolution of the
system is limited by the greater of the delay line resolution (<~40~fs in this setup) or probe pulse duration (80~fs in this setup).

The combination of pulse seed selection and delay line adjustment allows
roughly nine orders of magnitude temporal dynamic range. Fig. \ref{fig:sequence} exhibits
this dynamic range, showing the pre-plasma
expansion before the arrival of the main ultra-intense pulse (upper
sequence) and the hydrodynamic
response of the water jet over 10~\textgreek{m}s of evolution (lower sequence).

\section{Elimination of Plasma Self-Emission Noise}

\begin{figure*}
\includegraphics[width=6in]{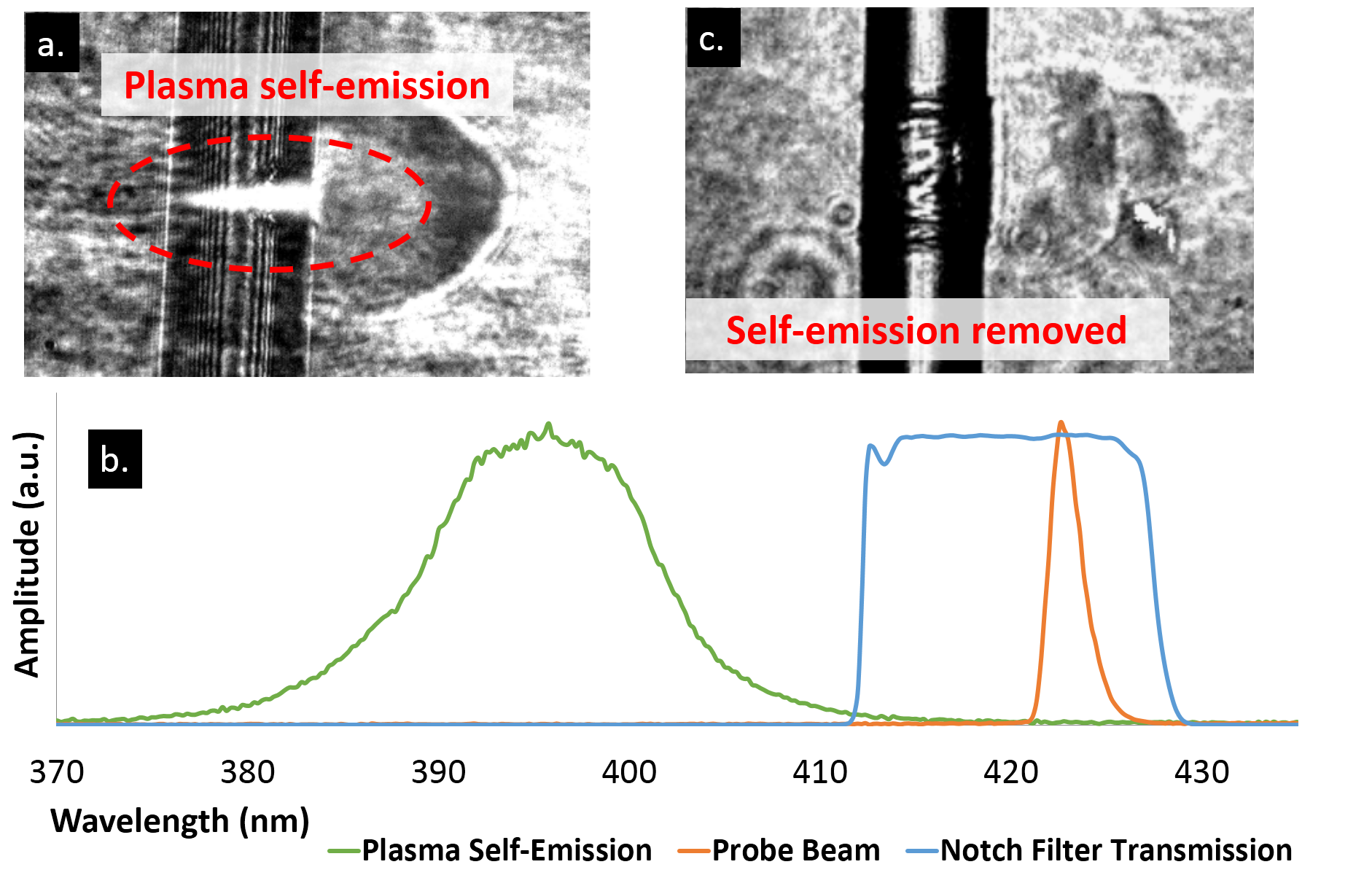}

\caption{\label{fig:notch} The deleterious effect of plasma self-emission
on the diagnostics is avoided by a frequency shift in the probe beam
and aggressive optical filtering. (a) Near the critical surface of
laser-plasma interaction, plasma self-emission at 400~nm saturates imagery. (b) To avoid this effect, the probe beam is
frequency shifted to 840~nm, amplified, then frequency doubled to
420~nm. At the imaging cameras, notch filters are used to
remove plasma self-emission at the pump's second harmonic,
400~nm. The notch filter transmission band includes
the entire probe beam spectrum and only the tail of the plasma self-emission
spectrum. (c) The effect of the frequency shift and optical notch
filters is to reveal previously obscured interaction areas.}
\end{figure*}

Irradiated by the pump beam at 800~nm, the target will naturally emit
light at the second harmonic (400~nm) \cite{von_der_linde_second_1992},
which is the source of the plasma self-emission that is the subject
of this subsection. Plasma self-emission complicates and frustrates
probe-beam interferometry and shadowgraphy because it obscures areas
of interest near the critical surface (See Fig. \ref{fig:notch}a).

The solution to this obfuscation problem, which we have implemented
in this setup, is to apply optical filtering that excludes plasma
self-emission while passing the probe beam. Even with specially selected
optical notch filters, this is only possible if the probe spectrum
sits outside of the plasma self-emission spectrum. However, because the pump and
probe share a common oscillator, the probe pulse naturally shares
the frequency of the pump harmonic. To circumvent this issue, the
frequency of the probe pulse is shifted in two steps.

Prior to amplification, the broadband 800~nm seed pulse of the probe
beam is is optically filtered using Schott RG850 glass, moving the central wavelength
to 840~nm by preferentially attenuating lower wavelengths. Next, a hard spectral cutoff is applied in the pulse stretcher,
by physically clipping lower wavelength regions of the beam during grating dispersion.
This frequency-shifted (and much attenuated) pulse is next passed
through a regenerative amplifier, which smooths the spectral profile,
and after re-compression the pulse remains centered at 840~nm. Finally,
the probe beam is frequency doubled to 420~nm via a BBO crystal (AR/AR coated, 300 \textgreek{m}m thick, Type I). During imaging, two optical notch filters (Semrock FF01-420/10-25;
10~nm bandwidth, and nominally 15~nm FWHM) select for the 420~nm probe
and reject the plasma self-emission at 400~nm (Figs. \ref{fig:notch}b
and \ref{fig:notch}c).

\section{Acquisitions and Analysis}

\begin{figure*}
\includegraphics[width=6in]{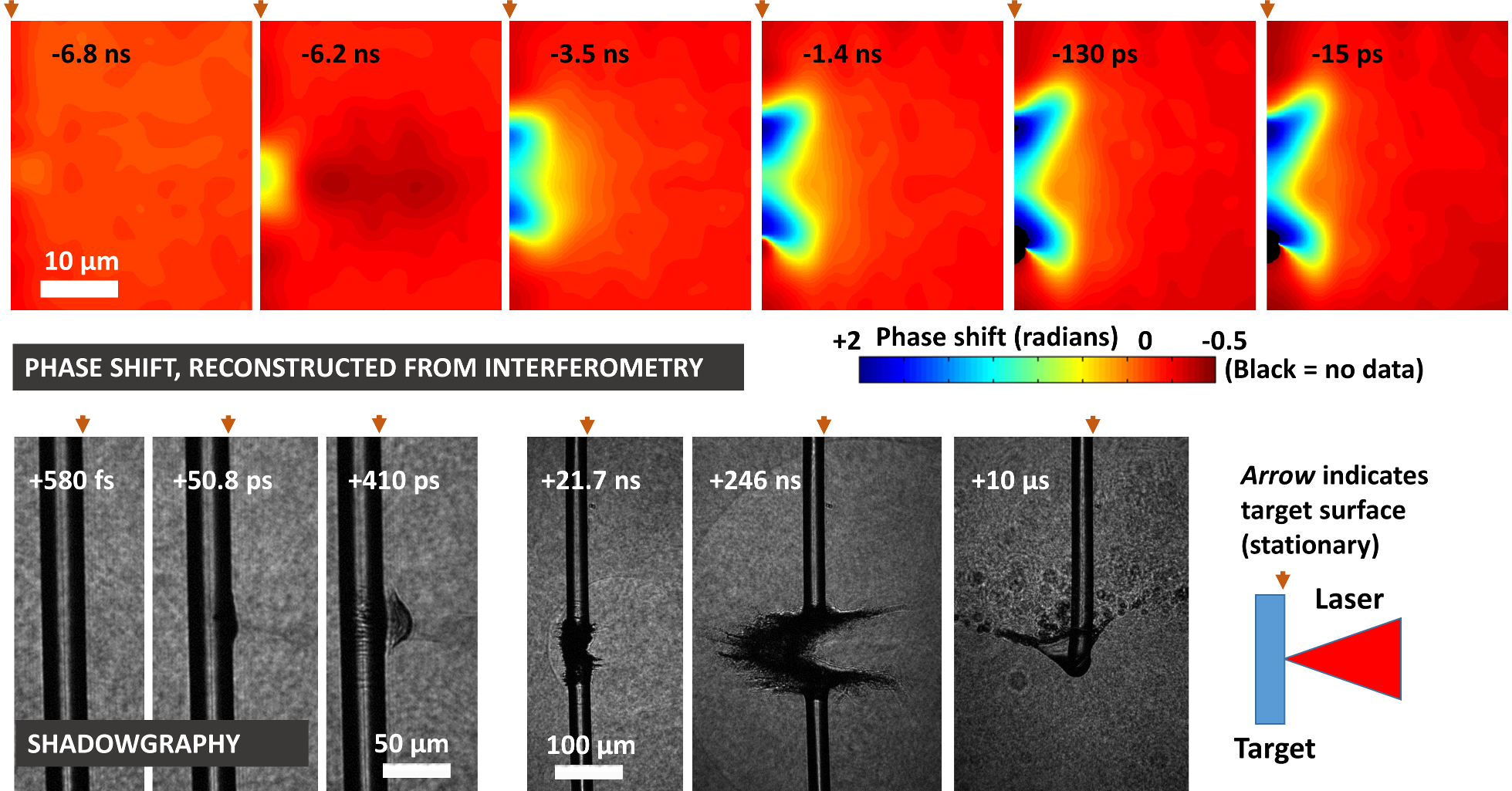}\caption{\label{fig:sequence}Precise probe pulse timing (<~40~fs resolution)
over a wide dynamic range (>~10~\textgreek{m}s) gives a detailed view
of pre-plasma evolution, intense laser-plasma interaction, and target
recovery. Upper sequence: Recovered phase shifts from interferometry show the in-vacuum development of pre-plasma nanoseconds before the main
interaction, consistent with known pre-pulse on this timescale.
Delay times relative to the main ultra-intense laser pulse are marked.
Interferometric reconstruction after the main pulse arrives (\ensuremath{\ge}
0~fs, not shown) fails due to steep phase gradients. Lower sequence:
Shadowgraphy shows timescales of in-air target evolution:
hydrodynamic reaction in picoseconds, expansion in nanoseconds, and recovery in microseconds.}
\end{figure*}

By varying probe beam delay time, an image sequence can be created
that shows the development of pre-plasma, the main ultra-intense pulse
interaction and the hydrodynamic recovery of the target (see Fig.
\ref{fig:sequence}). The image acquisitions are performed in the
following way. A sequence of delay times is chosen so as to acquire
many frames shortly before the main pulse interaction while keeping
a wide temporal view of the interaction. Once programmed, the sequence
is serially executed. To create a given delay, Pockels cell seed selection
and delay line position are adjusted automatically. For each delay,
one shadowgraphic frame and three interferometric frames are acquired
(corresponding to three positions of the piezoelectric mirror which
varies Michelson interferometer phase difference). Because stochastic
events can cause occasional extreme outliers shot-to-shot, 10 images
are acquired for each frame, corresponding to ten independent laser
shots. The image most similar to the average is incorporated as the
frame. 

\begin{figure*}
\includegraphics[width=6in]{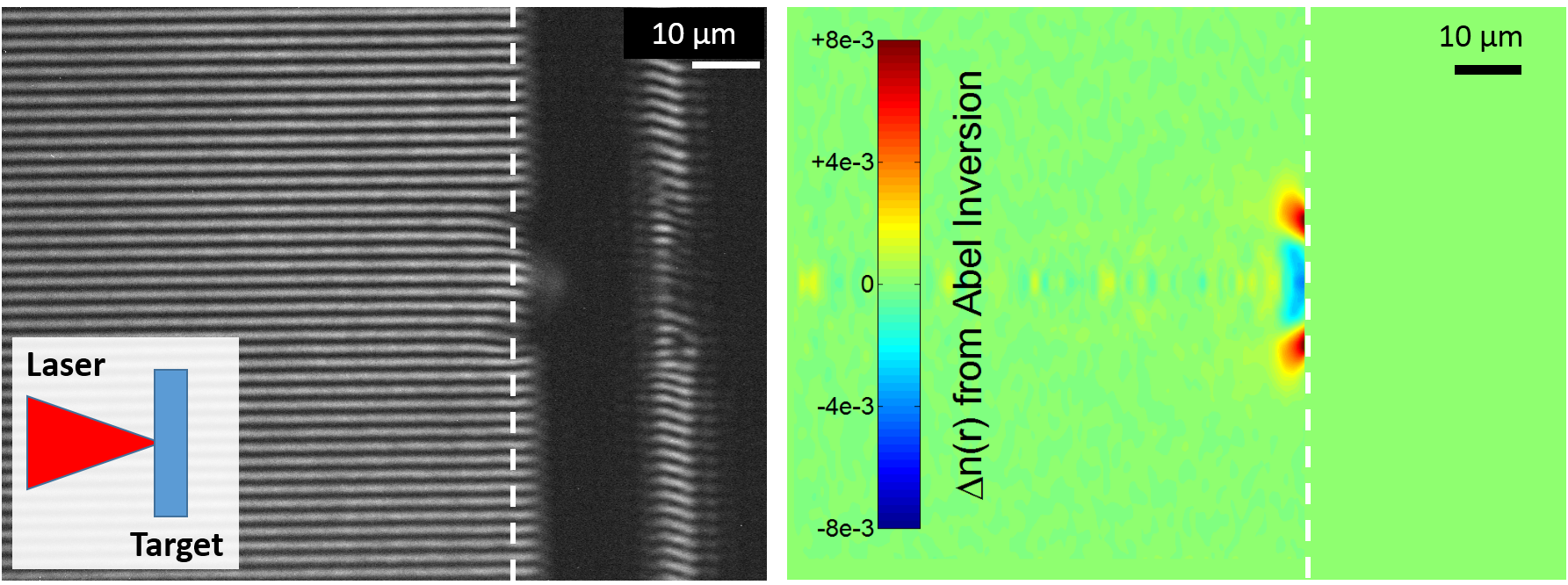}

\caption{\label{fig:abel} To aid in the reconstruction of phase dependent
imagery from fringe shifts, three separate 120\textdegree{} phase-shifted
(temporally sheared) interferograms are acquired for each probe delay.
Left, an interferogram is shown for -15~ps delay for an in-vacuum target. Its
corresponding phase image is reconstructed using the Speckle Phase
of Difference algorithm in IDEA software -- the vertical dashed line
represents the rightmost boundary of reliable phase reconstruction.
Abel inversion allows recovery of index of refraction and potentially
electron density. Right, recovered index of refraction shifts for the
interferogram at left are shown. The radial index change from negative
to positive suggests pre-plasma on-axis and ablated neutral material
off-axis.}
\end{figure*}

Interferograms must be analyzed to recover phase data. Phase shifts are reconstructed from
fringe shifts using the IDEA software \cite{hipp_digital_2004} Speckle
Phase of Difference algorithm (three-frame technique,
120\textdegree{} phase shift). Abel inversion
can be performed to recover the changes in index of refraction (see
Fig. \ref{fig:abel} for an experimental example). Abel inversion
analysis requires an added assumption of the experiment's radial symmetry,
about the laser axis. From the Abel inversion, electron densities
can potentially be inferred.

\section{Summary and Conclusions}

We have described a frequency-shifted, femtosecond-gated interferometric
setup and demonstrated its capability for a flowing water jet experiment
in which this target is irradiated by nanosecond-scale and picosecond-scale
pre-pulses in advance of the arrival of an ultra-intense (10\textsuperscript{18}~W/cm\textsuperscript{2}) pulse. The combination of a timing system
for observing evolution from femtoseconds to microseconds and elimination
of plasma self-emission noise light delivers improved shadowgraphy
and interferometry. Phase shift reconstruction from interferometry reveals
changes in the index of refraction near the target. This information
can be used to infer electron densities through Abel inversion. With
precise optical filtering and by frequency shifting the probe pulse
away from the self-emission frequencies, the problem of plasma self-emission
is avoided.

The diagnostic has been demonstrated to produce interferometric and
shadowgraphic image sequences of laser-matter interactions which show
nanosecond formation of pre-plasma, femtosecond interaction of the
ultra-intense main pulse, picosecond hydrodynamic expansion, and microsecond
recovery of the target. Fielding the diagnostic has already contributed
in two ways to our experimental understanding. First, tens-of-microseconds
target recovery time indicates that the experiment could be performed
at significantly higher than kHz repetition rates. Second, expectations
prior to the arrival of the main pulse of pre-plasma formation along
the laser axis with neutral material off-axis are confirmed. By providing
on-demand femtosecond resolution at arbitrary delay times, this diagnostic
will soon be used to track the effects of known femtosecond-duration
pre-pulses as they arrive on target picoseconds to nanoseconds prior
to the main pulse, and to improve laser-driven electron acceleration
and X-ray production techniques \cite{orban_backward-propagating_2014}.
Future improvements to the diagnostic may include additional views
of the interaction region or on-the-fly interferogram analysis for
fast feedback to the experimenter.

\begin{acknowledgments}
This research was sponsored by the Quantum and Non-Equilibrium Processes
Department of the Air Force Office of Scientific Research, under the
management of Dr. Enrique Parra, Program Manager. SF was supported
in part by the DOD HPCMP high performance computing internship program.
The authors thank Mario Manuel, University of Michigan, for insightful
discussions.
\end{acknowledgments}

\end{document}